\documentclass[times, letter]{mn2e}
\usepackage{natbib_jrm,graphicx,amssymb}

\bibpunct[, ]{(}{)}{;}{a}{}{,}

\def \aj {AJ}
\def \mnras {MNRAS}
\def \pasp {PASP}
\def \apj {ApJ}

\def \apjl {ApJL}
\def \aap {A\&A}
\def \nat {Nature}
\def \araa {ARAA}
\def \iaucirc {IAUC}

\def \kms {$\mathrm{km\;s^{-1}\;}$}
\def\lesssim{\mathrel{\hbox{\rlap{\hbox{\lower4pt\hbox{$\sim$}}}\hbox{$<$}}}}
\def\gtrsim{\mathrel{\hbox{\rlap{\hbox{\lower4pt\hbox{$\sim$}}}\hbox{$>$}}}}
\begin{document}
\title[Spectropolarimetry of the Type Ib/c SN 2005bf]{Spectropolarimetry of the Type Ib/c SN 2005bf\thanks{Based on observations made with ESO Telescopes at the Paranal Observatory, under programme 75.D-0213.}}
\author[Maund et al.]{
\parbox[t]{\textwidth}{\raggedright 
Justyn R. Maund$^{1}$\thanks{Email: jrm@astro.as.utexas.edu}, 
J. Craig Wheeler$^{1}$,
Ferdinando Patat$^{2}$, 
Dietrich Baade$^{2}$, 
Lifan Wang$^{3}$,
and Peter H\"{o}flich$^{4}$ }
\vspace*{6pt}\\
$^{1}$ Department of Astronomy and McDonald Observatory, 1 University Station C1400, University of Texas, Austin, Texas, 78712, U.S.A.\\
$^{2}$ ESO - European Organisation for Astronomical Research in the Southern Hemisphere, Karl-Schwarzschild-Str.\ 2, 85748 Garching\\ b.\ M\"unchen, Germany\\
$^{3}$ Department of Physics, Texas A\&M University, College Station, Texas 77843-4242, U.S.A.\\
$^{4}$ Department of Physics, Florida State University, Tallahassee, Florida 32306-4350, U.S.A.\\
}
\maketitle
\begin{abstract}
We present spectropolarimetric observations of the peculiar Type Ib/c
SN 2005bf, in MCG+00-27-005, from 3600-8550\AA.  The SN was observed
on 2005 April 30.9, 18 days after the first B-band light-curve maximum
and 6 days before the second B-band light-curve maximum.  The degree
of the Interstellar Polarization, determined from depolarized emission
lines in the spectrum, is found to be large with $p_{max}(ISP)=1.6\%$
and $\theta(ISP)=149$\fdg$7\pm4.0$, but this may be an upper limit on
the real value of the ISP.  After ISP subtraction, significant
polarization is observed over large wavelength regions, indicating a
significant degree of global asymmetry, $\gtrsim 10\%$.  Polarizations
of 3.5\% and 4\% are observed for absorption components of Ca II H\&K
and IR triplet, and 1.3\% for He I 5876\AA\ and Fe II.  On the $Q-U$
plane clear velocity-dependent loop structure is observed for the He I
5876\AA\ line, suggestive of departures from an axial symmetry and
possible clumping of the SN ejecta. Weak High Velocity components of
$\mathrm{H\alpha}$, $\mathrm{H\beta}$ and $\mathrm{H\gamma}$ are
observed, with velocities of $-15$ $000$\kms. The low degree of
polarization observed at H$\beta$ suggests that the polarization
observed for the other Balmer lines ($\sim 0.4\%$ above the background
polarization) may rather be due to blending of $\mathrm{H\alpha}$ and
$\mathrm{H\gamma}$ with polarized Si II and Fe II lines, respectively.
We suggest a model in which a jet of material, that is rich in
$\mathrm{^{56}Ni}$, has penetrated the C-O core, but not the He
mantle.  The jet axis is tilted with respect to the axis of the
photosphere.  This accounts for the lack of significant polarization
of O I 7774\AA, the delayed excitation and, hence, observability of He
I and, potentially, the varied geometries of He and Ca.
\end{abstract}

\section{Introduction}
\label{intro}
The presence of asymmetries in core-collapse supernovae (CCSNe) has
long been thought inherent to the nature of the explosion mechanism.
The history of observations of runaway O-stars and pulsars gave the
first direct evidence for such asymmetries
\citep{2005MNRAS.364...59D}.  Studies of the profiles of emission
lines in the late-time SNe spectra \citep{2005Sci...308.1284M} and the
morphologies of young remnants \citep{2006ApJ...645..283F} also
provided important clues.  The link between the SNe and the
highly-collimated Gamma Ray Burst (GRB) phenomenon
\citep{2006ARA&A..44..507W}, along with studies of the spherical
nature of Type IIP SNe for their use as distance indicators
\citep{hampi02,2001ApJ...553..861L}, has necessitated direct studies
of SN asymmetries at all stages of their evolution.\\
Spectropolarimetry, along with the new era of 8m-class telescopes, has
permitted the routine observation of CCSNe, inferring the asymmetries
directly from the polarization properties of these SNe.  While other
techniques, such as the modelling of emission line profiles, probe the
interaction between the SN ejecta and the circumstellar medium (CSM),
only spectropolarimetry provides a direct measure of the geometries of
SN ejecta and, more importantly at earlier epochs, the nature of the
geometry of the core-collapse mechanism itself.  The linear
polarization of light, induced by electron and line scattering, in
spherically symmetric atmospheres are completely cancelled, leading to
no net observed linear polarization.  In the presence of
irregularities or asymmetries, a net linear polarization is observed,
with the magnitude of the polarization and the polarization angle
related to the relative size of the asymmetry and the orientation of
the asymmetry on the sky
\citep{1982ApJ...263..902S,1984MNRAS.210..829M,1991A&A...246..481H}.\\
CCSNe are generally classified by the absence (Type I) or presence
(Type II) of hydrogen in their early spectra \citep{fili97}.  Type I
CCSNe are generally observed to have higher degrees of polarization,
suggesting higher degrees of asymmetries, than for Type II CCSNe at
early times.  This is inferred as the direct observation of the core
layers in Type I SNe, which are shielded by the hydrogen ejecta in
Type II SNe \citep{2001ApJ...550.1030W,2006Natur.440..505L}.  Once the
hydrogen layers of Type II SNe become optically thin, revealing the
inner core layers, the polarization is observed to increase
dramatically (see SN 2004dj, \citealt{2006Natur.440..505L}; SN 2001ig,
\citealt{my2001ig}).  Observations of the Type Ic SN 2002ap, which was
considered a possible host for a GRB misaligned with the line of
sight, showed it to have polarizations of $\gtrsim1\%$, corresponding
to asymmetries of $\sim20\%$
\citep{2002PASP..114.1333L,2003ApJ...592..457W}. Additionally, the
orientation of the observed polarization, through studies of the
Stokes $Q$ and $U$ parameters, showed that O and Fe had different
distributions within the SN ejecta \citep{2003ApJ...592..457W}.\\ SN
2005bf was discovered independently by Monard and the team of Li and
Moore \citep{2005IAUC.8507....1M} on 2005 Apr 6.72 UT.  Li and Moore
determined the position of SN 2005bf as
$\mathrm{\alpha_{2000}=10^{h}23^{m}57.27^{s}}$ and
$\mathrm{\delta_{2000}=-3\degr11'28.6''}$, $11.7''$ east and $32.6''$
south of the nucleus of the host galaxy MCG+00-27-005 (see
Fig. \ref{2005bf_pos}).  The heliocentric recessional velocity of the
host is given by
HyperLEDA\footnote{$\mathrm{http://leda.univ-lyon1.fr/}$} as
$v_{\odot}=5655$\kms, with the recessional velocity corrected for
infall to Virgo, $v_{\mathrm{Vir}}=5623$\kms, yielding a distance of
75Mpc (for $\mathrm{H_{0}=75 km\,s^{-1}\,Mpc^{-1}}$).\\ SN 2005bf was
a peculiar Type Ib/c SN.  It showed showed a double-peaked light curve
with a first maximum at $\sim16$ days post-explosion and a second,
brighter maximum at $\sim 40$ days post-explosion
\citep{2006ApJ...641.1039F}, assuming an explosion date of 2005 March
28 \citep{2005ApJ...633L..97T}.  At early epochs little or no He was
observed, giving some resemblance to a Type Ic SN
\citep{2005IAUC.8509....2M,2005IAUC.8509....3M}.  By the second
maximum, He I was observed in the spectrum
\citep{2005IAUC.8521....2W}, giving the appearance of a Type Ib SN
\citep{2005ApJ...633L..97T,2006ApJ...641.1039F}.
\citet{2005ApJ...633L..97T} and \citet{2006ApJ...641.1039F} invoke
``jets'' and ``holes'' to explain the observed behaviour.  If either
behaviour is involved, one would expect strong asymmetries to be
revealed by spectropolarimetry \citep{2001ApJ...550.1030W}.\\ Here we
present spectropolarimetry of SN 2005bf to elucidate the nature of the
explosion.  We find that neither published model accounts for the
spectropolarimetry and suggest a new configuration that may better
account for the observations.  The observation of SN 2005bf is
presented in Section \ref{obs}, with the results of these observations
presented in \S\ref{results}.  In \S\ref{analysis} these results are
analysed, and they are subsequently discussed in \S\ref{disc}.

\section{Observations}
\label{obs}
\begin{figure}
\caption{FORS1 1s V-band image of the location of SN 2005bf, relative
to its host galaxy MCG+00-27-005.  The SN is indicated by the
cross-hairs, and the arrow indicates the direction of the ISP
determined in Section \ref{isp}.}
\label{2005bf_pos}
\end{figure}\noindent
SN 2005bf was observed on 2005 Apr 30.9 (all times are UT), using the
European Southern Observatory (ESO) Very Large Telescope (VLT) Kueyen
Telescope with the Focal Reducer and Low Dispersion Spectrograph 1
(FORS1) instrument in the spectropolarimetric PMOS mode
\citep{1998Msngr..94....1A}.  These observations are summarised in
Table \ref{obstab}.  The FORS1 instrument was used with the standard
resolution collimator, providing a plate scale of
$\mathrm{0.2\arcsec\;px^{-1}}$.  The standard ``striped'' PMOS slit
mask was used, with slit width 1\arcsec and length 22\arcsec, and
throughout the entire set of observations the slits were kept at a
position angle $\mathrm{PA=0\degr}$.  Observations of SN 2005bf were
conducted with the ``super achromatic'' retarder plate positioned at 4
angles: 0\fdg0, 45\fdg0, 22\fdg5 and 67\fdg5.  The two beams, at each
retarder plate position, were dispersed using the G300V grism, which
provides a dispersion of 2.6\AA$\mathrm{\;px^{-1}}$ and spectral
resolution, measured from arc lamp calibration exposures, of 12.3\AA.
In this instance an order separation filter was not used leading to
some additional flux contamination at
$\mathrm{\lambda\gtrsim7000\AA}$.  The total wavelength coverage of
the data was $\mathrm{\sim 3600-8550\AA}$.  FORS1 uses a
$2048\times2048$ Tektronix CCD detector, and the observations were
conducted with gain $\mathrm{0.71e^{-}ADU^{-1}}$, with a readout noise
of 5.6$\mathrm{e^{-}}$.  The data were reduced in the standard manner
using IRAF\footnote{IRAF is distributed by the National Optical
Astronomy Observatories, which are operated by the Association of
Universities for Research in Astronomy, Inc., under cooperative
agreement with the National Science Foundation -
$\mathrm{http://iraf.noao.edu/}$} and a series of our own routines,
written in the Perl Data
Language\footnote{$\mathrm{http://pdl.perl.org/}$}.  The data were
corrected for bias and overscan.  A master unpolarized normalised flat
was constructed from flat observations acquired with the retarder
plate at each of the four angles.  The flat was applied to object
frames, and object spectra, for the ordinary and extraordinary rays at
each of the four retarder plate angles, were optimally extracted.
These spectra were wavelength calibrated through comparison with
observations of HeHgArCd arc lamps.  The Stokes parameters $Q$ and
$U$, the total polarization $p$ and the polarization angle $\theta$
were calculated in the standard manner \citep[see][]{forsman}, with
the data re-binned to 15\AA\ to improve the signal-to-noise (S/N) for
the individual Stokes parameters.  A correction was applied for the
wavelength-dependent chromatic zero angle offset.\\ The
spectropolarimetric calibration of the FORS1 instrument was checked by
observing, with 30s exposures at each retarder plate angle, the
spectropolarimetric standard Vela1 95, on 2005 Apr 30.98.  These data
were reduced in the same manner and found to be consistent with
previous measurements.  A single 300s observation of GD 153, with full
polarimetry optics and the retarder plate at 0\fdg0, was used to
provide a flux calibration and facilitate the removal of telluric
features from the observations of SN 2005bf. At each retarder plate
angle the average spectrum wide S/N $\sim 250$.  The data were
corrected for the heliocentric recessional velocity of the host
galaxy.  Statistical uncertainties of the measured Stokes parameters
were estimated using a Monte Carlo simulation of the FORS1
instruments, in a similar style to the model of
\citet{2006PASP..118..146P}.
\begin{table}
\caption{\label{obstab}  Spectropolarimetric ESO VLT FORS1 observations of SN 2005bf on 2005 Apr 30/May 1}
\begin{tabular}{lccc}
Object          & Avg.       & Retarder Plate &  Exp. time          \\
                &   Airmass  & Angles         &  (s)                \\
\hline
Vela 1 95$^{1}$ &   1.09     & 0,45,22.5, 67.5& $4\times30$         \\
SN 2005bf       &   1.13     &0,45,22.5, 67.5 & $4\times2\times1200$\\
GD153    $^{2}$ &   1.46     &   0            &    300              \\
\\
\end{tabular}
\\
$^{1}$ Polarized Standard\\
$^{2}$ Flux Standard\\
\end{table}

\section{Observational Results}
\label{results}
\begin{figure*}
\includegraphics[width=12cm,angle=270]{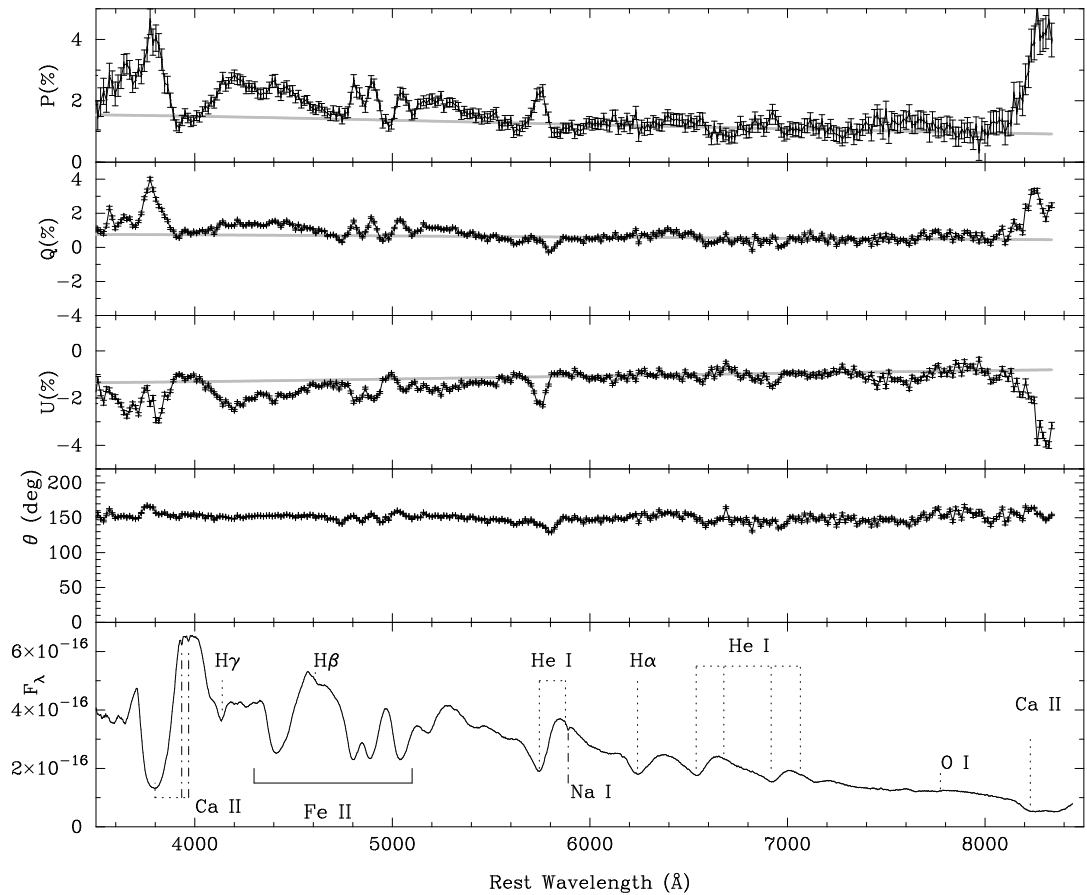}
\caption{Spectropolarimetric observation of SN 2005bf, from 2005 Apr
30.9.  The observation was conducted using ESO VLT FORS1.  The panels
show {\it (from top to bottom)} the polarization $p$, $Q$ Stokes
parameter, $U$ Stokes parameter, the polarization angle $\theta$ and
the total flux spectrum ($\mathrm{ergs\;s^{-1}\;cm^{-2}\AA^{-1}}$).
The wavelength scale of the data has been corrected for the
recessional velocity of the host galaxy.  The polarization parameters
have been re-binned to 15\AA, while the flux spectrum is at 2.6\AA\
$\mathrm{\;bin^{-1}}$.  The light smooth lines are best fits for a
Serkowski law-type Interstellar Polarization, for
$p_{max}(ISP)=1.6\%$, $\theta_{ISP}=149$\fdg$7$ and
$\lambda_{max}(ISP)=3000\mathrm{\AA}$.  Dotted lines, in the bottom
panel, indicate spectral features intrinsic to the SN, and dot-dashed
lines indicate lines arising from the ISM in the host galaxy.}
\label{2005bfspec}
\end{figure*}

The spectropolarimetric properties of SN 2005bf are shown as
Fig. \ref{2005bfspec}.  The epoch of the observations (JD2453492.5)
corresponds to 34 days post-explosion, and 18-days after the first
maximum and 6 days prior to the secondary, brighter visual maximum
\citep{2006ApJ...641.1039F} or 9-days prior to the second bolometric
maximum \citep{2007PASP..119..135P}.
\subsection{General Spectroscopic Properties}
\label{obsspc}
\citet{2005IAUC.8521....2W}, \citet{2005ApJ...631L.125A},
\citet{2006ApJ...641.1039F} and \citet{2007PASP..119..135P} have
previously presented discussion of the spectroscopic properties of SN
2005bf at various epochs.  Here we briefly summarise the properties of
this observation of SN 2005bf, to serve as an orientation for the
discussion to follow.\\ The flux spectrum of SN 2005bf (the bottom
panel of Fig. \ref{2005bfspec}) is dominated by a series of broad P
Cygni profiles, with several features attributable to He.  The
dominant He I feature is that of 5876\AA, with the minimum of the
absorption component at $-6\;700$\kms.  It is noted, however, that the
emission component of the He I 5876\AA\ P Cygni profile is not centred
on the rest wavelength, rather it is displaced blueward by $\sim
1\;500$\kms.  The emission components of the He I 5016,6678,7065\AA\
lines are similarly blue shifted, with similar absorption velocities
as the He I 5876\AA\ line.\\ At the blue extreme of the spectrum Ca II
H\&K are clearly visible as a single profile, and a P Cygni absorption
associated with the Ca II IR triplet is visible at the red extreme of
the spectrum.  The single Ca II H\&K absorption corresponds to an
average velocity of $-11\;700$\kms, and the IR triplet absorption
minimum corresponds to a velocity of $\sim -12\;500$\kms.\\ A series
of Fe II lines are observed as P Cygni profiles and identified,
following \citet{2005ApJ...631L.125A}, as: lines 4520, 4549, 4584,
4629\AA\ (multiplets 37 and 38) as a single P Cygni profile and 4924,
5015, 5169\AA\ (multiplet 42).  The absorption minima correspond to
expansion velocities of $-7\;700$\kms.\\ Despite classification as a
Type Ic SNe, \citet{2005IAUC.8521....2W} and
\citet{2007PASP..119..135P} have identified possible High Velocity
(HV) components of hydrogen $\mathrm{H\alpha}$, $\mathrm{H\beta}$ and
$\mathrm{H\gamma}$ at $-15\;000$\kms.  $\mathrm{H\beta}$ appears as
the only unambiguous Balmer feature, superimposed on the Fe II (37,38)
emission component.  $\mathrm{H\alpha}$ and $\mathrm{H\gamma}$ are
possibly blended with Si II 6355\AA\ and Fe II 4233\AA\ at the
photospheric expansion velocity (as defined by the velocities measured
for He and Fe II lines;
\citealt{2005IAUC.8521....2W,2005ApJ...631L.125A,2005ApJ...633L..97T,2006ApJ...641.1039F,2007PASP..119..135P}).
Higher Balmer features are not observed due to the strong Ca II H\&K
feature and fading response of FORS1 at bluer wavelengths.\\ Weak and
narrow interstellar absorptions of Ca II H\&K and Na I D are observed
at the recessional velocity of the host galaxy.  There are no
comparable absorptions for Milky Way interstellar absorption and,
given the resolution of the observations and the high recessional
velocity of the host galaxy, such components would be resolved if
present.  \citet{schleg98} give the fore-ground Galactic reddening at
$E(B-V)=0.01$.  The equivalent width of the Na I D line was measured,
through fitting a Gaussian line profile, as
$\mathrm{W_{\lambda}(Na\;ID)=0.7\pm0.1\AA}$.  Using the relations of
\citet{2003fthp.conf..200T}, this corresponds to a reddening of
$E(B-V)<0.37$.

\subsection{General Spectropolarimetric Properties}
\label{obspol}
\begin{figure}
\includegraphics[width=8cm,angle=270]{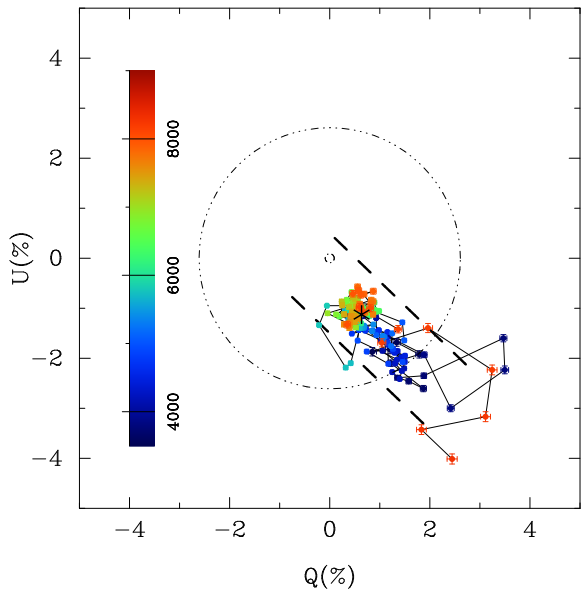}
\caption{Spectropolarimetric observation of 2005bf plotted on the
$Q-U$ plane.  The points are colour coded according to their
wavelengths following the scheme of the colour bar on the left hand
side of the figure, and the data have been re-binned to 30\AA.  The
data are distributed along a preferred axis, between the two heavy
dashed lines (which indicate the approximate extent of the data on the
$Q-U$ plane, and are parallel to the dominant axis uncorrected for the
ISP).  The limits on the ISP, as determined from reddening
considerations, are shown as circles for a){\it inner circle} Galactic
reddening $E(B-V)=0.01$ and $p<0.09\%$ and b){\it outer circle} host
reddening $E(B-V)=0.37$ and $p<3.3\%$.  The location of the determined
ISP component (see \S\ref{isp}), at 5500\AA, is indicated by
asterisk.}
\label{2005bfqured}
\end{figure}

The polarization properties of SN 2005bf are shown in the top four
panels of Fig. \ref{2005bfspec}.  The polarization $p$ shows a
significant gradient across the observed wavelength range, with higher
polarization at the blue end of the spectrum and a decreasing level of
polarization at longer wavelengths.\\ The polarization of the strong
lines observed in the flux spectrum exhibits ``inverted P-Cygni''
profiles, with the polarization peaking at absorption minimum and
depolarization over the emission component.  High levels of
polarization can be seen to be associated with Ca II H\&K (peaking at
$\sim 4.5\%$), Fe II lines, He I 5876\AA\ ($2.5\%$)and the absorption
of the Ca II IR triplet (peaking at $5\%$, although the feature is
only observed at a low S/N).  This is in contrast to the weak
polarization observed at $\mathrm{H\gamma}$.  The Fe II absorption at
4410\AA\ \citep{2003ApJ...592..457W,2006PASP..118..791B} shows only a
slight increase in $\Delta p \sim0.3\%$ compared to background levels
of polarization.  There is no significant polarization associated with
O I 7774\AA, above the background level of polarization.\\ The
polarization angle $\theta$ rotates as it passes over the absorption
features; changing by +20$\degr$ at the Ca II H\&K absorption and
-30$\degr$ for He I 5876\AA.  The polarization angle is, however,
generally constant at $\mathrm{\theta}=158$\fdg3$\pm0.1$ away from
conspicuous polarization associated with spectral lines.  The
consequence of the consistency of the polarization angle $\theta$ is
evident on Fig. \ref{2005bfqured}, which shows that the $Q$ and $U$
data fall along a dominant axis on the $Q-U$ plane.  The only
significant deviations from the dominant axis are across the
absorption and emission features associated with P Cygni profiles,
which have components along an axis orthogonal to the dominant axis on
the $Q-U$ plane.  The significant polarization ($\sim1\%$) observed
across depolarizing emission lines is suggestive that there is high
degree of non-intrinsic polarization, with the gradient across the
wavelength range being consistent with a \citet{1975ApJ...196..261S}
interstellar polarization (ISP) law.  The determination of the ISP is
presented in Section \ref{isp}.
\section{Analysis}
\label{analysis}
\subsection{Interstellar Polarization}
\label{isp}
\begin{figure}
\includegraphics[width=8cm,angle=270]{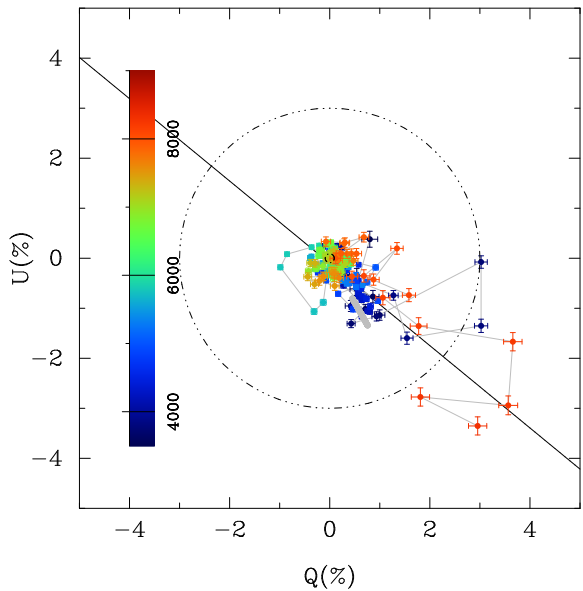}
\caption{Spectropolarimetric observation of SN 2005bf plotted on the
$Q-U$ plane, with the determined wavelength-dependent ISP subtracted.
The dominant axis of the data is indicated by the straight line.  The
points are colour coded according to their wavelengths following the
scheme of the colour bar on the left hand side of the figure, and the
data have been re-binned to 30\AA.  The wavelength dependent ISP is
indicated by the grey line.  The He I 5876\AA, Ca II H\&K and Ca II IR
Triplet are seen to significantly deviate, in the form of loops, from
the central concentration of data within $p < 0.45\%$ of the origin.}
\label{2005bf_qunored}
\end{figure}

The removal of the ISP is an important step in directly measuring the
intrinsic polarization properties of SNe from the observed data.
\citet{2003ApJ...591.1110W} discuss three considerations for
determining the ISP directly from SN observations: 1) Assume that the
emission lines are completely depolarized; 2) Assume the ISP is an
unvarying polarization component between different epochs in the SN
evolution; and 3) assume that observations of the SNe at the nebular
phase are intrinsically unpolarized due the low density of scattering
particles.  In the case of this observation of SN 2005bf the last two
points are invalid, since there is only one observation at one epoch
(so that a time invariant component cannot be identified) and the SN
is clearly not in the nebular phase.  The removal of the ISP from our
observation of SN 2005bf relies, therefore, on a number of assumptions
and being able to place useful limits of the size of the
ISP.\\\noindent Limits on the degree of polarization can be set by
considering the amount of reddening along the line of sight and the
relation between reddening and the degree of the ISP.
\citet{1975ApJ...196..261S} find that $p_{ISP}(\%)<9E(B-V)$, such that
reddening places a maximum limit on the value of the ISP.  For the
three reddenings determined in Section \ref{results}, we find the
$p_{ISP}<0.09\%$, for Galactic foreground reddening, and $3.3\%$, for
internal reddening in the host as measured by the Na I D line.  These
maximum limits are presented, on the $Q-U$ plane, on
Fig. \ref{2005bfqured}.  The majority of the reddening of SN 2005bf
arises in the host galaxy, as the only Na I D component observed is at
the redshift of the host.  In addition, in the catalogue of
\citet{2000AJ....119..923H}\footnote{Referenced using Vizier -
$\mathrm{http://vizier.u-strasbg.fr/}$} there are two polarized
Galactic stars, HD 90994 and 92886, which lie within $5\degr$ of SN
2005bf on the sky.  Both of these stars are at a Galactic Latitude of
$b=46\degr$ and distances $\mathrm{<460pc}$, and have polarizations
$<0.1\%$ suggesting only a small Milky Way contribution to the ISP.
The \citet{1975ApJ...196..261S} relation only provides an upper limits
on $p_{ISP}$, which only serves to provide a guide for more direct
techniques which aim to determine the ISP directly from the
observations of the SN itself.\\\noindent We note from
Fig. \ref{2005bfspec} that the emission features of Ca II H\&K, the Fe
II lines from 4900-5300\AA\ and the He I 5876\AA\ line display
significantly lower polarizations than either the associated
absorption features \citep{1984MNRAS.210..829M} or the continuum in
the immediate vicinity.  This suggests that there is a high level of
intrinsic polarization to the SN, but that, even for features that are
expected to be completely depolarized, the ISP is $\sim 1\%$.  Because
of the relatively high level of SN polarization, with respect to the
ISP, a single vectorial subtraction of constant values of $Q_{ISP}$
and $U_{ISP}$ at every wavelength is inappropriate.  Instead, it is
required that the ISP be calculated at each wavelength, according to a
\citet{1975ApJ...196..261S} law, where:
\begin{equation}
p ( \lambda ) =p_{max} \mathrm{exp} \left(-K{\mathrm{ln}}^{2}\left(\lambda_{max}/ \lambda \right)\right)
\label{serkfit}
\end{equation}
where \citet{1992ApJ...386..562W} give:
\begin{equation}
K=0.01+1.66\lambda_{max}(\mathrm{\mu m})
\end{equation}
  The calculation of the corresponding Stokes parameters
$Q_{ISP}(\lambda)$ and $U_{ISP}(\lambda)$ requires the assumption of a
constant polarization angle and, hence, that $Q_{ISP}(\lambda)$ and
$U_{ISP}(\lambda)$ are in the same ratio across the wavelength range.
In addition, the ISP can be located on the $Q-U$ plane by demanding
that it lie along the dominant axis of the observed data
\citep{2003ApJ...591.1110W}.  In the case of SN 2005bf such a dominant
axis is observed, and the ISP can be located at either end of
distribution of the data on the $Q-U$ plane.  Given the high density
of depolarizing resonance lines at the blue end of the spectrum, it is
expected that the ISP is located on the $Q-U$ plane closest to the
bluest data along the dominant axis \citep{2001ApJ...556..302H}.  The
polarization angle associated with the depolarizing emission lines is
measured as $\theta_{ISP}=149$\fdg$7\pm4.0$ (see
Fig. \ref{2005bfqured}).  This result is surprising, since there is an
expectation that the ISP should be aligned with the spiral arms of the
host galaxy, as shown as Fig. \ref{2005bf_pos}
\citep{1987MNRAS.224..299S}.  SN 2005bf is, however, significantly
displaced from the nearest spiral arm and the case, therefore, for the
polarization angle of the ISP is not expected be as straightforward as
the cases of SN 2001ig \citep{my2001ig} or SN 2006X
\citep{nando2006X}.  The validity of the assumption used here to
calculate the ISP is discussed in \S\ref{subsecisp}.\\ A simultaneous
$\chi^{2}$-fit of a \citeauthor{1975ApJ...196..261S} ISP law (of the
functional form given in Eqn. \ref{serkfit}), for the parameters
$p_{max}$ and $\lambda_{max}$, was conducted using the narrow
wavelength regions corresponding to the minimum degrees of
polarization, at the peaks of the emission lines in the flux spectrum.
The $\chi^{2}$ minimization was conducted for the polarization $p$ and
the two Stokes parameters $Q$ and $U$.  In this way the value of the
polarization angle of the ISP could also be tested, since $\chi^{2}$
should be minimized for the same values $p_{max}$ and $\lambda_{max}$
for independent fits to $p$, $Q$ and $U$, if $\theta_{ISP}$ is
correct.  Statistically significant values of $\chi^{2}$ and
simultaneous minima in the $\lambda_{max}-p_{max}$ plane were only
achieved for values of $\theta_{ISP}=149$\fdg$7\pm4.0$.  This
demonstrates that our value of the polarization angle of the ISP is
consistent with our assumption that the Fe II lines are depolarized.
There is a certain degree of degeneracy between $p_{max}$ and
$\lambda_{max}$, but the rising degree of polarization observed at the
emission lines suggested that the polarization was increasing blueward
and that $\lambda_{max}$ was located blueward of the observed data.
The best-fit values were $p_{max}=1.6\pm0.2\%$ ($99\%$ confidence) at
$\lambda_{max}=3000\mathrm{\AA}$. \\ The ISP and the Stokes parameters
$Q_{ISP}(\lambda)$ and $U_{ISP}(\lambda)$ are shown on
Fig. \ref{2005bfspec}.  The subtraction of the wavelength-dependent
ISP, shown on the $Q-U$ plane as Fig. \ref{2005bf_qunored}, contracts
the data about the origin, except for the polarization associated with
certain lines.  The degree of the ISP across the B and V photometric
bands ($\sim 1.5\%$) conversely limits the amount of reddening along
the line of sight to $E(B-V)>0.17$.\\ The values of $\lambda_{max}$
determined here is substantially bluer than that measured by
\citet{1975ApJ...196..261S} for Galactic stars.  \citet{nando2006X}
found a similar value of $\lambda_{max}$ was applicable for the ISP
for SN 2006X; however, in that case the wavelength dependence of the
ISP did not conform to a standard Galactic
\citeauthor{1975ApJ...196..261S}-type law or a value of $K$ consistent
with \citet{1992ApJ...386..562W}.

\subsection{Intrinsic Polarization}
\label{intpol}
After the subtraction of the ISP, significant polarization is still
observed across the spectrum, particularly associated with spectral
lines.  Given the shallow wavelength-dependence of Eqn. \ref{serkfit}
the ISP can, potentially, remove some intrinsic polarization from the
data but cannot remove polarization associated with spectral features
that vary widely over short wavelength ranges.  Specific polarization
features are discussed at length in the following sections.  There
appears to be a degree of ``continuum'' polarization, for regions of
the spectrum over which the Stokes parameters hardly vary.  This is
the case for data in the range $\mathrm{6000<\lambda<8000\AA}$, where
the average polarization is $0.45\%$, but shows no apparent
correlation with features in the flux spectrum.  The bulk of the data,
on the $Q-U$ plane (see Fig. \ref{2005bf_qunored}), lies within
$p\approx0.45\%$ of the origin, with the obvious exceptions of He I
and Ca II lines.  This degree of polarization implies asymmetries of
the order $\gtrsim 10\%$ \citep{1991A&A...246..481H}.

\subsection{He I 5876\AA}
\label{heiline}
\begin{figure}
\includegraphics[width=8cm,angle=270]{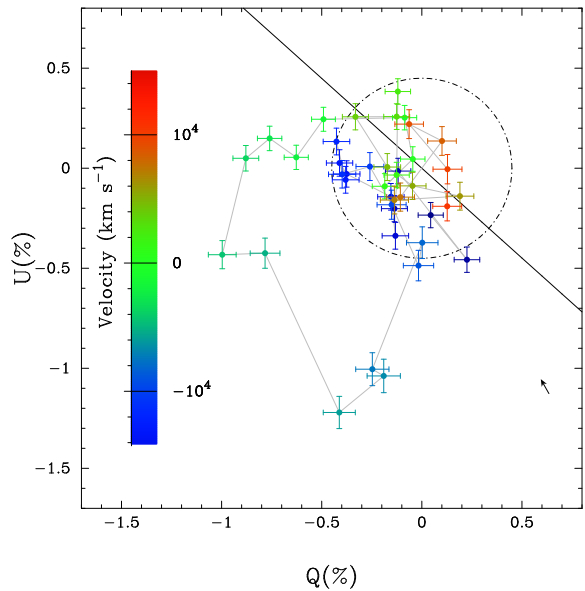}
\caption{The $Q$ and $U$ Stokes parameters around the He I 5876\AA\
line.  The data have been re-binned to 15\AA\ (or 765\kms\ at
5876\AA).  The points are colour coded by the observed velocity (\kms)
relative to the rest wavelength of the line.  The dominant axis,
determined for the entire data set as shown in
Fig. \ref{2005bf_qunored}, is shown as the straight black line.  The
arrow indicates the degree of ISP which, due to wavelength dependence,
shows the degree of ISP from blue to red wavelengths.  The dot-dashed
circle shows the region of ``continuum'' polarization $p=0.45\%$.}
\label{2005bf_he}
\end{figure}
As discussed previously, the He I 5876\AA\ line shows a significant
degree of polarization, rising from a continuum level of 0.2\% to
1.3\% at the absorption minimum.  The emission component, after ISP
subtraction, has no polarization, as assumed in the calculation of the
ISP presented in Sec. \ref{isp}.  The increase in polarization,
anti-correlated with the flux spectrum of the absorption line, is
matched by a rotation in the polarization angle corresponding to
$0\degr$ at the continuum level to $~50\degr$ at the line centre.
This corresponds to the production of a loop, as seen on
Figs. \ref{2005bfqured}, \ref{2005bf_qunored} and, specifically,
\ref{2005bf_he}.\\ The effect of the ISP, and in particular its
orientation, stretches the loop in the direction of the ISP angle and
away from the origin of the Q-U plane.  While the subtraction of the
ISP causes the loop to be contracted, it is still clear that the loop
structure, on the $Q-U$ plane, is intrinsic to the SN itself.\\ The
loop structure, shown on Fig. \ref{2005bf_he}, lies on only one side
of the dominant axis, and the emission component can be seen to lie at
the origin of the $Q-U$ plane, since it is measured to be completely
depolarized.  The bulk of the loop structure is formed by wavelengths
blueward of the rest wavelength of this line.\\ In this case, a simple
straight-line fit to data on the $Q-U$ plane
\citep[see][]{2006ApJ...653..490W} would yield only a poor fit, since
it is clear that the structure is a loop.  There is no rotational
transformation on the $Q-U$ plane, to rotate the parameters to
``dominant'' and ``orthogonal'' axes, that can remove the loop
\citep{my2001ig}.

\subsection{Ca II H\&K and IR Triplet}
\label{cahkirtriplet}
The highest degrees of polarization, across the observed spectral
range, are measured for the Ca II H\&K and IR triplet as $2.9\pm0.7\%$
and $4\pm1\%$, respectively.  On the $Q-U$ plane, as plotted in
Fig. \ref{2005bf_qunored}, the Stokes parameters across these two
lines are observed to occupy approximately the same area with the same
orientation relative to the origin.\\ A relatively high degree of
uncertainty on these values is caused by low levels of signal-to-noise
at these wavelengths, due to the limited response of FORS1 at both the
blue and red extremes of the data.  Even so, the degrees of
polarization associated with these features are still significantly
higher than for any other features in the spectrum.\\ Whereas the He I
5876\AA\ line rotated by 50\degr, as discussed in Section
\ref{heiline}, a different amount of rotation of the polarization
angle is observed across these features (from $60$ through $85\degr$)
is observed; although in the case of the Ca II IR triplet the data is
incomplete as the emission component lies redward of the spectral
range of the observation.  Furthermore, the $Q-U$ data associated with
these features show that they are approximately aligned with the
determined ISP.  This leads to some concern over the veracity of the
measured value of the polarization angle across these features.  The
amount by which the polarization angle rotates across the lines should
not be affected, since that is a relative change.  Again, this problem
is most likely due to uncertainties in the Stokes parameters induced
by the low signal-to-noise ratio of the data at these wavelengths.
These effects would also cause the Stokes $Q$ and $U$ parameters to
deviate from the clear loop structure observed for such lines as He I
5876\AA.  If the Stokes parameters across these two lines are assumed
to be correct then the polarization angle measured across these two
lines are: 1) consistent for the two Ca lines, suggesting production
of both these lines by Ca ions with the same distribution within the
ejecta; and 2) that Ca has a significantly different distribution
within the ejecta compared to He.
\subsection{HV HI $\mathrm{H\alpha}$,  $\mathrm{H\beta}$ and  $\mathrm{H\gamma}$}
\label{highvelocityhi}
\begin{figure}
\includegraphics[width=8cm,angle=270]{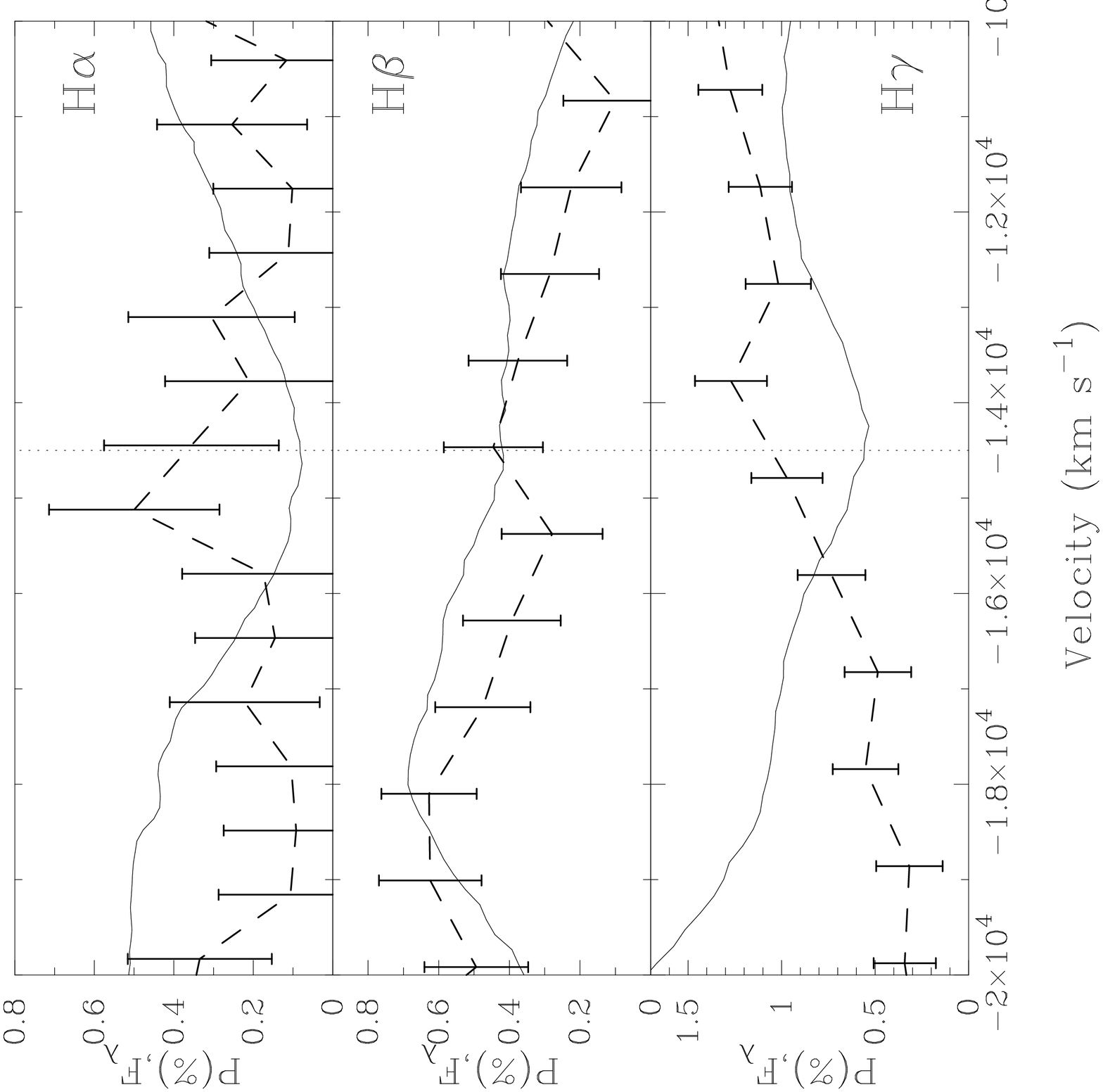}
\caption{The polarization across the high-velocity HI Balmer
absorption lines, as a function of velocity.  The flux spectrum,
arbitrarily scaled is shown as the solid black line, with polarization
measurements $p$ shown as points with error bars and the dashed lines.
The approximate central velocity for the absorption minimums of all of
the lines is indicated by the dotted line.}
\label{2005bf_balmer}
\end{figure}

As discussed in Section \ref{obsspc}, a series of absorption features
in the spectrum can be identified as $\mathrm{H\alpha}$,
$\mathrm{H\beta}$ and $\mathrm{H\gamma}$ at high velocities.
\citet{2005ApJ...631L.125A}, \citet{2005ApJ...633L..97T} and
\citet{2007PASP..119..135P} debate, however, the relative
contributions of H and other possibly blended species, such as Si II
and Fe II. Blending with other lines, with different polarization
properties, would cause differences in the observed polarizations for
each of the Balmer lines.  Spectropolarimetry can, therefore, be used
to assess the importance of blending, when considering particular line
identifications.\\
Small increases in the degree of polarization, across the Balmer features, are
shown as Fig. \ref{2005bf_balmer}.  The peaks in polarization are seen
to disagree with the absorption minima of the line profiles in the
flux spectrum for $\mathrm{H\alpha}$ and $\mathrm{H\gamma}$, even
though the minima for the three lines have similar velocities to
within 1 bin (or 765\kms) in the flux spectrum.\\ There is no
detectable loop structure associated with the HV HI features, which is
to be expected given that the polarization associated with these
features is low relative to the degree of the continuum polarization
either side of each of these lines, in wavelength space.  Instead, the
principal measurable signature of these lines is the total
polarization itself.
The smallest degree of polarization for these possible HV features is
observed for $\mathrm{H\beta}$, where the increase in polarization
over the background polarization is $0.1\pm0.1\%$.  The low degree of
polarization might have two possible causes: 1) the superposition of
the absorption on the depolarizing Fe II (multiplet 37,38) emission
component; and 2) the low optical depth which would give rise to such
weak absorption features in the flux spectrum. This is at odds with
higher degrees of polarization associated with $\mathrm{H\alpha}$ and
$\mathrm{H\gamma}$.  As discussed above, $\mathrm{H\beta}$ is the only
Balmer line not associated with possible absorption features due to
other species, leading to $\mathrm{H\beta}$ being effectively the
``cleanest'' Balmer absorption.\\ If homologous expansion is assumed,
then the HV H I components are exterior to the slower moving Fe II
lines, which implies that the depolarized photons of Fe II are
repolarized purely by $\mathrm{H\beta}$.  As presented in
Sec. \ref{obsspc}, $\mathrm{H\alpha}$ and $\mathrm{H\gamma}$ may be
contaminated by Si II 6355\AA\ and Fe II 4233\AA\ moving at only the
photospheric velocity.  Certainly in the case of the Fe II line there
are other comparison lines, discussed in Sec. \ref{felines}, which do
show significant polarization associated with their absorption
features.  It is more likely, perhaps, that the polarization signature
observed for $\mathrm{H\alpha}$ and $\mathrm{H\gamma}$ arises from
blends of the Si and Fe lines.  The polarization of the
$\mathrm{H\gamma}$ line is identical to other Fe II lines, in the
degree of polarization and the amount through which it rotates across
the absorption profile.  This indicates that Fe II 4233\AA\ is likely
to be responsible for most of the observed polarization for this
feature.  The degree of polarization intrinsic to the H I lines
themselves can, therefore, be limited by the amount observed for
$\mathrm{H\beta}$ to $p(\mathrm{H I})\lesssim 0.2\%$.  The observation
of any polarization associated with the HV Balmer series is
surprising, since the weakness of these lines suggests low densities
in the HV shells which would lead to only a small amount of
polarization actually associated with these lines.
\subsection{Fe lines}
\label{felines}
The difference in the polarization properties of the Fe II lines of
multiplets 37,38; and 42 can be viewed as primarily due to line
blending effects.  The close proximity of the Fe II lines of
multiplets 37,38 (in wavelength space) leads to the production of a
single P Cygni profile, but results in only a small increase
($\sim0.3\%$) in the local level of polarization.  The flux at the
absorptions associated with the redder lines of these two multiplets
are dominated by the depolarized flux from the emission of the bluer
lines.\\ For multiplet 42 there is sufficient separation of the lines
to yield resolved components in the flux spectrum, with resolved peaks
in the polarization spectrum.  If the degree of flux associated with
the absorption of the redder lines is significant, relative to the
amount of depolarized flux from the emission components of the two
bluest lines, there would be a significant intrinsic polarization
observed at the peaks of the emission components.  This would,
therefore, imply that the ISP determined by the assumption of
depolarization associated with these Fe lines is, in fact, an upper
limit of the degree of ISP.  The lack of further observations at
additional epochs, however, limits the discussion of just how
inappropriate this assumption, used to calculate the ISP, actually
is.\\ Given the significant degree of polarization associated with the
absorption components of the lines of Fe II multiplets 37,38 and 42,
it is to be expected that significant degrees would also be associated
with other Fe II lines.  \citet{2005ApJ...631L.125A} identify the
presence of Fe II $\lambda4233$ (of multiplet 27).  Given the measured
velocities of the redder Fe II lines, as 7700\kms, this implies that
the absorption of this Fe II line would be coincident with HV
$\mathrm{H\gamma}$ at $15\;000$\kms.  The increase in polarization
associated with the absorption feature at $\mathrm{\sim 4130\AA}$ is,
therefore, more likely to be due to Fe II absorption.  Furthermore,
there is possible blending of Fe II 5169\AA\ with Fe III 5156\AA\
which cannot be resolved.\\ It is noted that the maximum degrees of
polarization associated with Fe II lines are approximately similar
($\sim 1-1.5\%$), and that the Stokes Q and U parameters and the
polarization angles are also similar.  The Fe lines demonstrate a
larger degree of rotation in the polarization angle across the
absorptions profiles (from 20 to 80\degr, from the blue wing to the
absorption minimum, after correction for the ISP) than observed for
either Ca and He.  This suggests that the distribution of Fe is
different to that of both He and Ca in the SN ejecta.

\section{Discussion}
\label{disc}
A number of studies have been conducted of the photometric and
spectroscopic evolution of SN 2005bf.  In this respect, the amount of
spectropolarimetry data available for this SN is extremely low, but it
is important.\\ The low degree of intrinsic ``continuum''
polarization, $<0.45\%$, implies asymmetries of only $10\%$.
\citet{1991A&A...246..481H} shows that the observed polarization for
an oblate ellipsoid can be reduced by inclining it to the observer.
This may, indeed, be the case with SN 2005bf; there are, in any case,
obvious asymmetries in SN 2005bf.  A polarization of 0.45\% is larger
than the values of the continuum polarization reported by
\citet{2003ApJ...592..457W} for SN 2002ap at earlier epochs.  While
the photosphere itself may not be highly asymmetric, it is clear from
the polarization signatures associated with spectral lines that the
ejecta within the line forming region are highly asymmetric
\citep{1984MNRAS.210..829M} and, importantly, highly stratified.
Comparison of all the different species giving rise to lines with
strong polarization signatures shows they must occupy different parts
of the ejecta to give rise to different polarization angles.\\
\subsection{Loops on the Q-U plane}
 The presence of loop structure on the $Q-U$ plane for SN 2005bf is
quite similar to that observed for SN 2002ap by
\citet{2003ApJ...592..457W} The He I 5876\AA\ line of SN 2005bf may be
produced by the same mechanism that \citeauthor{2003ApJ...592..457W}
suggested for the O I 7774\AA\ line in SN 2002ap: the line forming
region and the continuum forming region beneath do not share the same
axial symmetry and can be characterised by two different polarization
angles.  The varying contribution of scattered flux from the blue edge
of the absorption to the absorption minimum leads to different ratios
of the polarized flux from the two regions and, hence, varying
polarization angle, leading to a loop on the $Q-U$ plane.  In the case
of a SN, this can arise from the absorption line forming region and
the photosphere not having the same shape.\\ Loops on the $Q-U$ plane,
associated with the absorption components of P Cygni profiles, were
identified by \citet{1988MNRAS.231..695C} in observations of SN 1987A.
A loop was observed for the Ca II HV component of the Type Ia SN
2001el and a series of models, including ellipsoidal shells and
clumped ejecta, were found to approximately reproduce such a feature
\citep{2003ApJ...593..788K}.  \citet{2006astro.ph.12244H} identified
loop structure in Type IIn SNe, and characterised the CSM as a series
of ellipsoidal shells with varying orientations.  \citet{my2001ig}
identified loop-like features in spectropolarimetry of the Type IIb SN
2001ig, which were interpreted as the presence of a significant
non-bipolar component of the ejecta \citep{2001AIPC..586..459H}.  As
\citet{2003ApJ...593..788K} indicate, however, the determination of
the structure of the ejecta from single loops in the $Q-U$ plane is
extremely difficult, requiring the computational sampling of a large
amount of parameter space.  While it is not clear how the loop of the
He I 5876\AA\ line should be interpreted, it is indicative of varying
symmetries in the SN ejecta rather than one major axial symmetry.\\ In
SN 2005bf we also observe large structures on the $Q-U$ plane for Ca
II H\&K and the IR triplet.  While the latter loop is incomplete and
the data for both features suffer from poor S/N, there is a
significant polarization associated with these lines.  The
polarization angles of these lines are at a different orientation than
the He I 5876\AA\ line, suggesting a different distribution in the
ejecta.  The flux spectrum also suggests this property, with the
absorption minima of both Ca lines occurring at substantially higher
velocities than the photospheric velocity determined from He I and Fe
II lines.  \citet{my2001ig} observe similar behaviour in the Ca II IR
triplet feature of SN 2001ig, where it does not share the same
orientation with the rest of the data from each particular epoch.
While \citeauthor{my2001ig} suggested some possible link with the
CSM-ejecta interaction, it is clear that here the Ca II component is
moving much slower than the HV Balmer lines and, unlike SN 2001el
\citep{2003ApJ...591.1110W}, Ca II is not observed as two separable
photospheric and HV components.  Again, we suggest some caution in
interpreting the polarization angle of the Ca II lines due to possible
correlation with the determined ISP.\\
\subsection{HV Balmer Components}
The HV components of the Balmer series that we observe here are
consistent with the expectations that thin shells of hydrogen may
remain on the massive progenitors of this type of SN.  HV components
are more commonly observed for the Ca II IR triplet in Type Ia SNe,
but they are also observed in CCSN types as they are produced by
approximately the same physics \citep{2007astro.ph..3468C}.
Spectropolarimetry of these features is rare, and the most direct
comparison is made with Ca II HV features in Type Ia SNe.
\citet{2003ApJ...591.1110W} observed polarizations of 0.7\% for SN
2001el, inferring a mass of $\mathrm{0.004M_\odot}$ in the HV
component.  \citet{2006astro.ph..9405C} did not, however, see any
polarization of that feature for the very similar SN 2004S.
\citet{nando2006X} have observed an HV component polarization of $\sim
1.1\%$ in SN 2006X, again a similar event to SN 2001el.  The upper
limit on the intrinsic polarization of the H I features of SN 2005bf
is substantially lower, therefore, than the polarizations measured for
HV features in Type Ia SNe.  The low upper limit of the amount of
polarization intrinsic to H I itself and the lack of any significant
deviation of the polarization angle from the background values make it
unlikely that the observed polarization is coming from a geometrically
distinct plane as \citet{2003ApJ...591.1110W} observed for Ca II in SN
2001el.\\
\subsection{SN 2005bf vs. SN 2002ap}
The opportunities to conduct spectropolarimetry of CCSNe are
increasing, however it is still too soon to have a complete set of
observations for all of the diverse types of CCSNe.
\citet{2005ApJ...631L.125A} state that SN 2005bf is a close relative
of the peculiar Type Ib SN 1999ex; but SN 2005bf is, to the best of
our knowledge, the only example of this type of SN with
spectropolarimetry.\\ The object with the most comparable
spectropolarimetry data set is SN 2002ap, which was observed by
\citet{2003ApJ...592..457W}, \citet{2002PASP..114.1333L} and
\citet{2002ApJ...580L..39K}.  SN 2005bf is similar to SN 2002ap in
that the Stokes parameters for the bulk of the data are concentrated
around the origin of the $Q-U$ plane.  Between 3
\citep{2003ApJ...592..457W} and 25 days \citep{2002PASP..114.1333L}
after V maximum, SN 2002ap shows lower levels of polarization ($\sim
0.3\%$) across large wavelength regions than this particular epoch of
SN 2005bf.  The only feature of SN 2002ap that has a distinct and
significantly different polarization is O I 7774\AA ; this line is not
observed to be significantly polarized in SN 2005bf.  Similarly, at 6
days prior maximum the polarization of SN 2002ap was dominated by the
O I 7774\AA\ line, with most of the spectrum at $p\lesssim 0.25\%$.
\citet{2003ApJ...592..457W} observed that the ``continuum''
polarization of SN 2002ap increased with age, become less concentrated
around the origin of the $Q-U$ plane.\\ For SN 2005bf we observe that
there are a number of features that are significantly more polarized
than the continuum polarization, most notably the Ca II lines and He I
5876\AA ; but also that the Stokes parameters of the data at the
``continuum'' are less concentrated about the origin of the $Q-U$
plane (within a radius of $p=0.45\%$) than \citet{2003ApJ...592..457W}
observed for SN 2002ap at early times.  A re-reduction of the entire
Wang et al. data set was conducted, to test whether the scatter of the
data on the $Q-U$ plane might have been due to a numerical effect in
our new data reduction routines.  The result of this additional test
was a complete set of Stokes parameters for all of the observations of
SN 2002ap which were completely consistent with the results presented
by \citeauthor{2003ApJ...592..457W}.  This demonstrates that our new
routines would have been able to easily resolve the structure measured
for SN 2002ap, and the scatter observed for the data of SN 2005bf is,
in fact, real and not due to the new routines.\\
\subsection{Estimate of the ISP}
\label{subsecisp}
There are a number of limitations to the conclusions that can be drawn
about the geometry of SN 2005bf, given that there is only an
observation at a single epoch.  There were multiple observations of SN
2002ap by \citet{2003ApJ...592..457W}, \citet{2002PASP..114.1333L} and
\citet{2002ApJ...580L..39K}.  The benefit from having data from
multiple epochs is clear, especially by being able to follow the
evolution of particular features.\\ Additional observations could
provide a direct measure of the ISP, by determining the non-varying
polarization component, due to the ISP, present at each epoch.  The
determination of the ISP presented here is based on an assumption that
emission lines are completely depolarized.  Given the degree of
potential blending of the Fe II lines, discussed in Sec \ref{felines},
it is possible that the emission lines are {\it not} completely
depolarized, in which case the determined ISP overestimates the real
value of the ISP \citep{1997PASP..109..489T}.\\ If the degree of the
ISP is lower than determined here, then the degree of the measured
continuum polarization would be higher leading to a higher degree of
inferred asymmetry.  The slight wavelength-dependence of the ISP,
however, would lead to less contraction of the data about the origin
of the $Q-U$ plane; since continuum polarization would be higher at
the blue end of the spectrum causing the data to be farther from the
origin of the $Q-U$ plane than the data at the red end of the
spectrum.  The measured Stokes parameters of the spectral lines would,
however, remain almost unchanged, since the polarization varies over
these spectral lines on significantly smaller wavelength scales than
the ISP.  The polarization measured for the spectral lines and the
conclusions drawn from them are, therefore, very real.  There is
concern about the measured polarization angle of the Ca features and
the similarity of these angles with the ISP; further
spectropolarimetry of SN 2005bf at other epochs would have revealed if
this is, indeed, merely coincidence.  If the other extreme of the ISP
is considered, whereby all of the polarization is intrinsic to the SN,
this would only imply asymmetries of the order $20\%$.\\
\subsection{Comparison with Two-Component Explosion Models}
\begin{figure}
\includegraphics[width=8cm,angle=0]{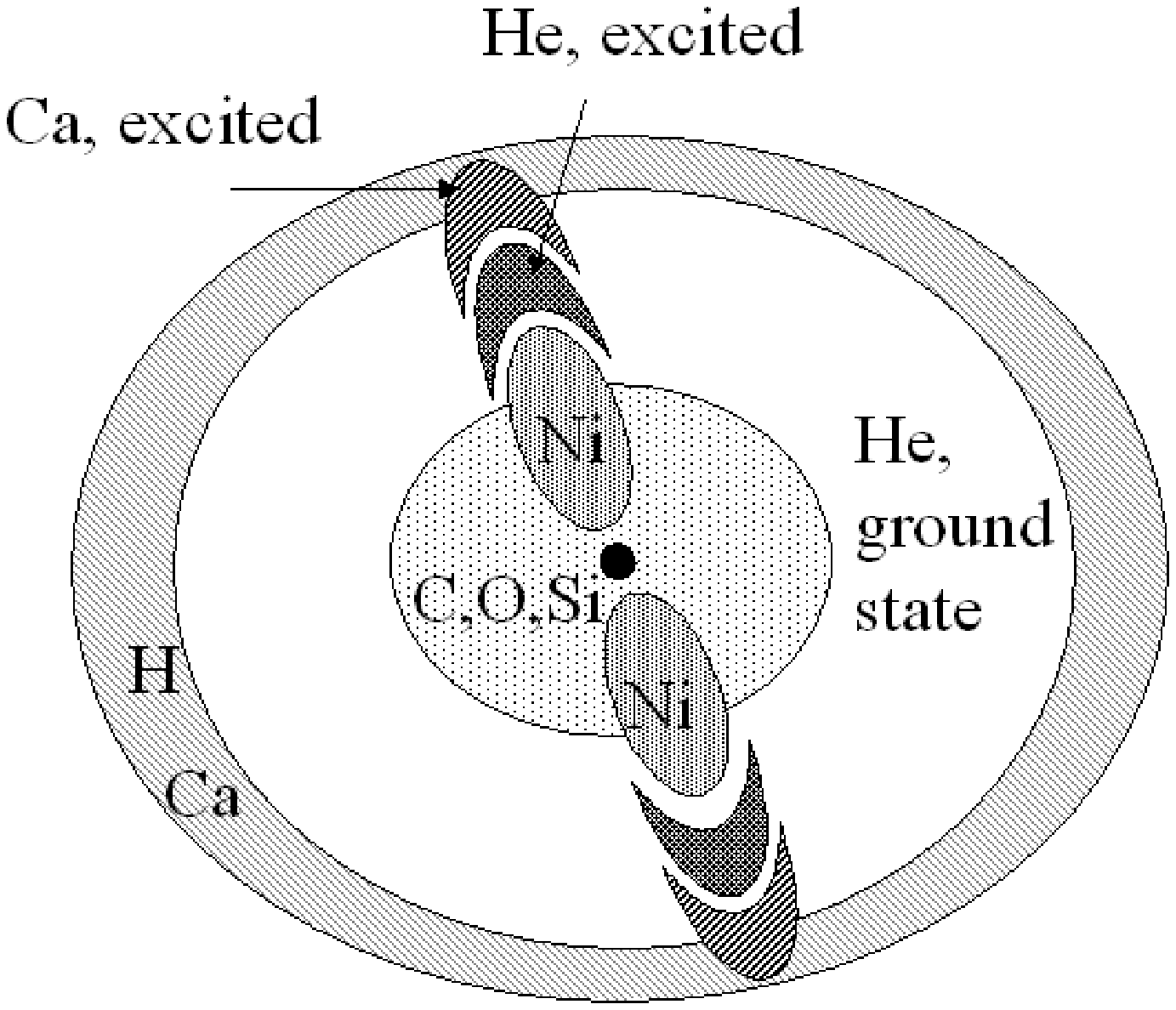}
\caption{A schematic of the structure of SN 2005bf, based on the
spectropolarimetry presented here and the models of
\citet{2005ApJ...633L..97T} and \citet{2006ApJ...641.1039F}.  The
observed polarization of He and Ca lines are due to asymmetric
excitation of these elements, due to asymmetric distributions of Ni
and, hence, gamma-ray deposition.  The continuum polarization arises
from the asymmetry of the photosphere.  The Ni jet axis is misaligned
compared to the axis of the photosphere.  Loops on the $Q-U$ plane for
He and Ca lines are produced by the different orientations of
asymmetries of these elements and the photosphere.  Core-material and
products of nucleosynthesis are shielded by the photosphere.  The
photosphere is immediately exterior to the C-O core, but is located
inside the He-envelope.  He exterior to the photosphere is unexcited.}
\label{2005bf_schematic}
\end{figure}

\citet{2005ApJ...633L..97T} and \citet{2006ApJ...641.1039F} presented
two distinct models to explain the spectroscopic evolution of SN
2005bf.  \citet{2006ApJ...641.1039F} presented a two-component model
with the first, polar, component dominating the spectra and light
curve at earlier epochs with the signature of a Type Ic SN, followed
by a lower velocity, more isotropic component containing most of the
He. \citet{2005ApJ...633L..97T} suggest that the emergence of He in
the spectrum was due to growth of ``holes'' in the ejecta, as it
expanded, permitting gamma rays to be deposited into the He envelope
leading to the excitation of He at later epochs.\\ Some polarization
of the He absorption feature may arise from the recombination front of
He just outside the photosphere, in this case the inferred asymmetry
would be consistent with that of the photosphere.  It is clear,
however, that the observed polarization of the He lines, and the
inferred asymmetry, is dominated by the excited He component that is
not coupled to the asymmetry of the photosphere.\\ It is clear,
however, from studies such as \citet{2005ApJ...633L..97T} and
\citet{2006ApJ...641.1039F} that SN 2005bf was not, at earlier epochs,
a ``classical'' Type Ic SN \citep[see][]{wheelersn}.  As mentioned
above, the weak O I 7774\AA\ line with no discernible associated
polarization is in stark contrast to Type Ic SNe such as SN 2002ap.
This suggests that the C-O core is still shielded by the photosphere,
which is located within the He envelope.  The weakness of the line and
the lack of polarization all indicate that the line is formed by
primordial O I in the He envelope.\\ The absence of peculiarly strong
iron features, with strong polarization signatures and substantially
different polarization angles compared to He and Ca, seems to suggest
that jet structure, originating from the core, is not present.  This
is in agreement with the lack of a highly polarized O I feature, that
would be expected if core material were brought to the photosphere by
a jet.  In addition, the orientation of the polarization angles of the
He I and Ca II features, along with the presence of low polarization
HV H I components, suggest the explosion of an He star.  It seems
unlikely, therefore, that a jet penetrated the surface of the star.
We suggest that the excitation of He I and Ca II is more likely due to
an asymmetric Ni distribution (perhaps due to a jet that stalled
inside the progenitor) that only at later epochs, due to decreasing
optical depth with the expansion of the SN, deposited energy directly
into the He envelope to produce He features and cause the transition
from a Type Ic to a Type Ib SN.  The spectropolarimetry data,
therefore, favours the model proposed by \citet{2005ApJ...633L..97T},
but with the specific interpretation that the ``holes'' are due to
partial penetration of the He envelope by a Ni-rich jet, reminiscent
of \citet{2006ApJ...641.1039F}.\\ This model is presented as
Fig. \ref{2005bf_schematic}.\\ Additional spectropolarimetry
observations at earlier epochs, in comparison with the observation at
this epoch, could have potentially directly distinguished between
these two models.  The presence of a jet, following the model of
\citet{2006ApJ...641.1039F}, would have had polarization
characteristics clearly distinct from that observed at this epoch.
Indeed, given the highly asymmetric nature of a jet, the high level of
polarization at earlier times may have followed similar behaviour as
that observed for the Type Ic SN 1997X up to 15 days post-explosion
\citep{2001ApJ...550.1030W}.\\
\section{Conclusions}
\label{conc}
A single epoch of spectropolarimetry of SN 2005bf, on 2005 Apr 30.9
and 9 days prior to second B light curve maximum, show significant
polarization.  There is a strong polarization component due to the
intervening interstellar medium in the host, corresponding to a
Serkowski law with $p_{max}=1.6\%$, $\lambda_{max}=3000\AA$ with
$\theta_{ISP}=149$\fdg$7\pm4.0$.  This limits the reddening towards SN
2005bf to $E(B-V)>0.17$. Possible blending effects of multiple Fe II
lines may cause this estimate to be an upper limit on the ISP.\\ After
subtraction of the ISP component, SN 2005bf is revealed to still be
strongly polarized in the continuum consistent with a physical
asymmetry of $\sim10\%$.\\ Even larger polarization and asymmetry is
correlated with spectral features.  The Ca II H\&K and IR triplet
absorption features show the highest degrees of polarization.  The
polarization angles, and the rotation of the polarization angle across
the line profiles, are the same.  Caution is recommended about
interpreting the absolute offset of the polarization angles of these
lines, since these lines were observed at the limits of the FORS1
response function.\\ Significant polarization is observed across the
He I 5876\AA\ line, peaking at 1.1\% above the background
polarization.  The rotation of the polarization angle across the He I
feature yielded a loop on the $Q-U$ plane, but the polarization angles
observed for He I 5876\AA\ were not aligned with those observed for
the absorptions of the two Ca lines.  The High Velocity H I lines are
observed, with $\mathrm{H\beta}$ being the least blended. Significant
increases in polarization across the profiles of $\mathrm{H\alpha}$
and $\mathrm{H\gamma}$ are suggested as most likely being due to
blends with the polarized absorption components of Si II 6355\AA\ and
Fe II 4233\AA, moving at the slower photospheric velocity.  The slight
detection of any polarization at HV $\mathrm{H\beta}$, superimposed on
a depolarizing Fe II line emission component, is proposed as evidence
that the features due to hydrogen are not significantly polarized
themselves to within $p(H I)\lesssim 0.2\%$.  This is most likely due
to the HV component being a low mass shell of H I and, hence, with low
optical depth.\\ Spectral features of Fe II and O I are observed to be
neither particularly strong nor excessively polarized, suggesting they
arose in the He envelope and were not asymmetrically deposited by a
jet penetrating the envelope.  The polarization of He I and Ca II
absorption features is due to increased deposition of gamma-rays from
Ni, due to the decrease in the optical depth to these gamma-rays with
the expansion of the SN ejecta.  A ``tilted-jet'' model may account
for the array of polarization features.
\section*{Acknowledgements}
The authors are grateful to the European Southern Observatory for the generous allocation of observing time. They especially thank the staff of the Paranal Observatory for their competent and never-tiring support of this project in service mode.  The research of JRM and JCW is supported in part by NSF grant AST-0406740 and NASA grant NNG04GL00G.
\bibliographystyle{mn2e}

\end{document}